\documentclass[superscriptaddress,twocolumn,showpacs,fleqn]{revtex4}

\usepackage{amsmath}
\usepackage{graphicx}

\renewcommand{\vec}{\mathbf}

\def\fig#1{Fig.\,\ref{#1}}

\def\eq#1{(\ref{#1})}

\def\fig#1{Fig.\,\ref{#1}}

\def\i{{\rm i}}

\def\oH{{\rm H}}\def\ot{{\rm t}}\def\op{{\rm p}}\def\ovp{\mathbf{p}}
\def\opx{{\rm p}_{\rm x}}\def\opy{{\rm p}_{\rm y}}\def\opz{{\rm p}_{\rm z}}
\def\ox{{\rm x}}\def\oy{{\rm y}}\def\oz{{\rm z}}
\def\orabs{{\rm r}}
\def\opeff{\mbox{{\footnotesize$\Pi$}}}

\def\sH{H}\def\st{t}\def\svp{\mbox{{\bfseries\it p}}}
\def\svp{\boldsymbol{p}}
\def\spx{p_{x}}\def\spy{p_{y}}\def\spz{p_{z}}
\def\sx{x}\def\sy{y}\def\sz{z}
\def\spr{p_{\varrho}}\def\sr{\varrho}\def\srabs{r}
\def\speff{\pi}

\begin{document}

\title{Dynamical characteristics of Rydberg electrons released by a weak electric field}
\author{Elias Diesen}
\affiliation{Max-Planck-Institut f{\"u}r Physik komplexer Systeme,
N\"othnitzer Stra{\ss}e 38, 01187 Dresden, Germany}
 
\author{Ulf Saalmann}
\affiliation{Max-Planck-Institut f{\"u}r Physik komplexer Systeme,
N\"othnitzer Stra{\ss}e 38, 01187 Dresden, Germany}
 
\author{Martin Richter}
\affiliation{Institut f{\"u}r Kernphysik, Goethe-Universit{\"a}t,
Max-von-Laue-Stra{\ss}e 1, 60438 Frankfurt am Main, Germany}
 
\author{Maksim Kunitski}
\affiliation{Institut f{\"u}r Kernphysik, Goethe-Universit{\"a}t,
Max-von-Laue-Stra{\ss}e 1, 60438 Frankfurt am Main, Germany}
 
\author{Reinhard D{\"o}rner}
\affiliation{Institut f{\"u}r Kernphysik, Goethe-Universit{\"a}t,
Max-von-Laue-Stra{\ss}e 1, 60438 Frankfurt am Main, Germany}

\author{Jan M. Rost}
\affiliation{Max-Planck-Institut f{\"u}r Physik komplexer Systeme,
N\"othnitzer Stra{\ss}e 38, 01187 Dresden, Germany}
\affiliation{PULSE Institute, Stanford University and SLAC National Accelerator Laboratory,
2575 Sand Hill Road, Menlo Park, California 94025, USA}

\begin{abstract}\noindent
The dynamics of ultra-slow electrons in the combined potential of an ionic core and a static electric field is discussed. 
With state-of-the-art detection it is possible to create such electrons through strong intense-field photo-absorption and to detect them via high-resolution time-of-flight spectroscopy despite their very low kinetic energy. 
The characteristic feature of their momentum spectrum, which emerges at the same position for different laser orientations, is derived and could be revealed experimentally with an energy resolution of the order of 1\,meV.
\end{abstract}

\maketitle\noindent
Extracting excited electrons from an atom or molecule, even a plasma with a constant electric field is an established technique which can reveal the minimal (Rydberg)-excitation by tracking the electron yield as a function of applied field strength \cite{ga05a}.
In the time domain, very small extraction fields of about $F\,{=}\,1\ldots10$\,V/cm (with $5.142\,\mbox{V/cm}=10^{-9}$ atomic units) and high, pulsed Rydberg excitation lead to intricate electron dynamics despite the fact that a hydrogenic problem in an electric field is separable (e.g., in semi-parabolic coordinates). Clearly, electrons which are capable to escape under such conditions must be highly excited and this is achieved in the experiment by a preceding excitation with a short intense laser pulse \cite{nugo+08}.
Interestingly, the details of this excitation are irrelevant for the dynamics described here.

The relevant questions of this problem are: Can one describe the dynamics classically or semi-classically? We will demonstrate, in comparison to experiment, that a classical description is very accurate and we will explain why this is the case.
A second question regards the momentum spectrum of these electrons: Is it structured or smooth? We will demonstrate that it is characterized by a main peak, offset from zero by a well defined momentum, in very good agreement with experiment.
The dependence of the peak position on the field strength $F$ is deduced analytically and compared to experimental results without any free parameters.

The kind of Stark dynamics discussed here with a focus on the differential momentum distribution of ionized electrons has not been investigated before, since ionization of Rydberg states in static or pulsed weak electric fields has mainly served the purpose to extract details of specific quantum states  ranging up to principal quantum number $n\,{\approx}\,100$ by either tunneling or over-barrier ionization  (ZEKE spectroscopy) \cite{ga05a,lano96,sc98}.  Individual classical trajectories and their contributions to the electron yield, on the other hand,  have been investigated both in the context of astrophysics, where the same Stark Hamiltonian arises from a combined gravitational and constant driving field \cite{laru11}, and in an atomic setting. For the latter, the inclusion of phases to account for interference effects of different pathways to a detector \cite{koos84, bo98a,bole+03} even lead beyond a classical treatment.
 
The results presented here are also  relevant for all experiments using electric-field extraction techniques, in particular the COLTRIMS \cite{ulmo+03} technique which is nowadays used routinely in many reaction microscopes (REMI) world wide. At very low energies, the Stark dynamics described here will modify the results expected in such setups. Moreover, it adds another interesting feature to the low-energy kinetic-energy spectrum of electrons ionized by an intense laser pulse, a process which has received a lot of attention \cite{kasa+12,kasa+12a,wuya+12,guha+13,mome+14,bego+14} since the discovery of the so called low-energy structure  \cite{blca+09,quli+09}.

To understand the phenomena under discussion, it is sufficient to consider a hydrogenic problem whose dynamics is governed by the Hamiltonian (atomic units are used unless stated otherwise and bold letters describe vectors)
\begin{equation}\label{eq:ham}
{\cal H} = \oH+\vec{\vec{r}}\,{\cdot}\,\vec{f}(\ot)\cos\omega \ot 
\end{equation}
with the Stark Hamiltonian
\begin{equation}\label{eq:ham1}
\oH = \frac{\ovp^{2}}{2}-\frac{Q}{\orabs} - F\oz
,\quad\orabs=\sqrt{\ox^{2}{+}\oy^{2}{+}\oz^{2}}
\end{equation}
where the ionic charge $Q$ is taken for the rest of the paper to be $Q =1$ (but could assume also an effective non-integer value) and a constant electric field of strength $F$ points in the negative $\oz$-direction. 
The second term in \eq{eq:ham} describes the interaction with the laser pulse of (strong but otherwise arbitrary) envelope
$\vec{f}(\ot)$, linearly polarized along or perpendicular to the direction of the extraction field along the $\oz$-direction in $\oH$. It is this combination of fields which leads to the ``zero-energy'' structure
observed in the strong field photo ionization spectrum with REMIs \cite{duca+13}. However, the role of the strong laser pulse is merely to populate the Rydberg states which serve as an initial condition for the unusual Stark dynamics we are going to discuss. Any other means to create the Rydberg population would lead to the same result if the Rydberg atoms are placed in a tiny electric field of the order of 10\,V/cm: The photo-electron spectrum exhibits a peak {\em not} at zero energy but slightly upwards at a field-corrected photo-electron momentum (defined in Eq.\,\eqref{eq:peff} below) of $\opeff^{*} = -0.6 F^{1/4}$, with the asterisk indicating the peak value.

\begin{figure}[t]
\centerline{\includegraphics[width=\columnwidth]{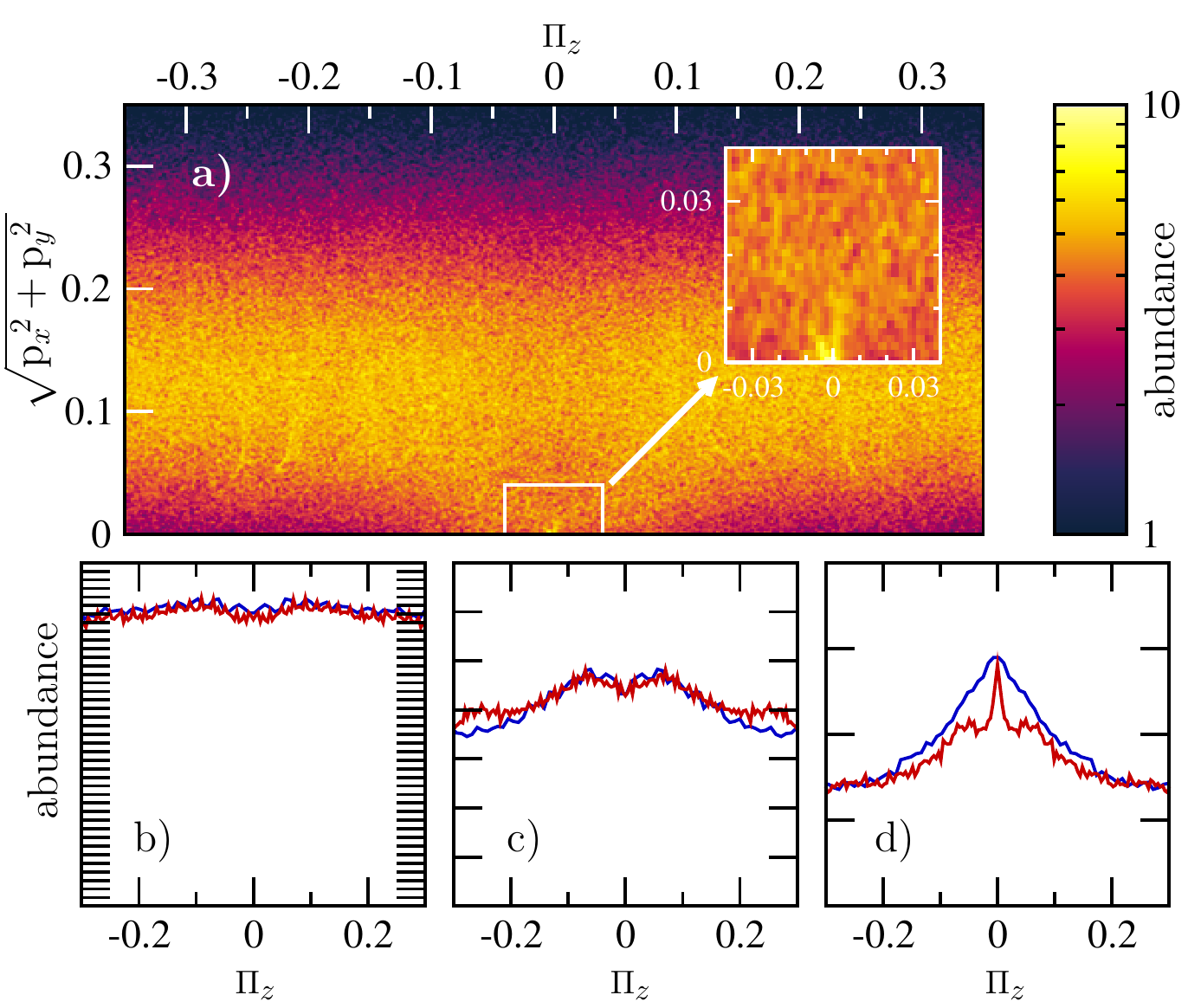}}
\caption{a) 2D momentum map of photo-electrons from Ar atoms ionized
with a laser pulse $\vec{f}(\ot)=\vec{e}_{z}f(\ot/T)\cos(\omega\ot)$ with $T=64$\,fs and $\omega=0.3875$\,eV ($\lambda=3.2$\,$\mu$m) and an extraction field $F=1.5$\,V/cm along the $\oz$-axis. \\
Lower panels: Spectra as obtained by integrating the 2D distributions over b) $\op_{\perp}=0\ldots0.35$,
 c) $\op_{\perp}=0.02\ldots0.06$, and  d) $\op_{\perp}=0.0\ldots0.02$, respectively.
Calculations (red lines) are compared to experimental data (blue, \cite{duca+13}).
Note that these spectra are symmetrized with respect to $\opeff_{\parallel}$.
}
\label{fig:3200nm}
\end{figure}%
Note that we have assumed $\ell_{z}=0$ (equivalent to the magnetic quantum number $m=0$) and thus ignore the dynamics in the cyclic azimuthal angle. 
This reduction is possible as the laser-induced dynamics occurs on a length scale (given by the quiver amplitude, typically a few nm), which is much smaller than the barrier distance (typically $\mu$m) of the Stark Hamiltonian \eq{eq:ham1}, and confirmed by the excellent agreement with full 3D simulations.
Since the momentum of the electron increases under $\oH$ as $\opz\propto \sqrt{\oz}$ or $\opz\propto\ot$, with distance $\oz$ from the nucleus or time $\ot$, respectively, its value depends on where it is measured. To avoid such a dependence, the experimental time-of-flight $\ot_{\rm of}$ is used to construct an effective momentum 
\begin{equation}\label{eq:peff}
\opeff \equiv \opz(\ot_{\rm of}) - F\,\ot_{\rm of},
\end{equation}
which eliminates the effect of the constant electric field asymptotically. 
In other words, $\opeff$ does not depend on the position of the detector as long as it is placed sufficiently far beyond the barrier. This is a convenient procedure since experimentally the time of flight $\ot_{\rm of}$ and the detector distance $\oz_\mathrm{d}$ are known. Consequently, one obtains $\opeff=\oz_\mathrm{d}/\ot_{\rm of} -F\,\ot_{\rm of}/2$, which is equivalent to \eq{eq:peff}.
In quantum mechanics, the momentum variable $\opeff$ can be elegantly obtained directly through the interaction representation with $\oH_{\ot}={\cal U}\oH {\cal U}^{\dagger}$ where 
${\cal U} = \exp(-\i F\oz\ot)$. Similarly, in classical mechanics, a canonical transformation with the generating function 
${\cal F}_{2}= (\opz{+}\ot)\bar z$ leads to new canonical variables $\bar{\op}_{\oz}=\opeff$ and $\bar\oz=\oz$.

We note that in all experimental studies using COLTRIMS reaction microscopes $\opeff$ in Eq.\,\eqref{eq:peff} is referred to simply as the momentum component $\opz$ of the electron parallel to the field direction of the spectrometer \cite{ulmo+03}. 
The underlying tacit assumption is that the motion of the electron can be split into an interaction region, where the extraction field can be neglected, and an asymptotic region, into which the electron is launched  with an initial momentum $\opz$.
It is this separation which breaks down for the situation discussed here.

In \fig{fig:3200nm} we compare classical simulation according to $\cal H$ given in \eq{eq:ham} with experimental results \cite{duca+13} obtained for argon atoms in a strong few-cycle pulse with $\lambda\,{=}\,3200$\,nm.
Those experiments were the first to report a ``zero-peak'', i.\,e.\ an extremely narrow contribution of near-zero momentum electrons \cite{duca+13}, subsequently confirmed and attributed to Rydberg electrons \cite{wole+14}. 
We use standard tunnel-ionization probabilities \cite{shgo+09,kasa+12a} and propagate electrons according to Newton equations of motion for about 10$^{7}$\,trajectories with the Hamiltonian \eq{eq:ham}, i.\,e., in the attractive Coulomb potential, the driving laser pulse $\vec{f}(t)\cos\omega t$ and the extraction field $F$. 
Hence, the calculations comprise the formation of Rydberg electrons \cite{nugo+08} as well the subsequent Stark dynamics.
The experimental results \cite{duca+13} shown in \fig{fig:3200nm}b--d are symmetrized in $\opeff$ with respect to $\opeff=0$ and show a double peak very close to $\opeff=0$. As one can see, the likewise symmetrized classical spectra agree extremely well with the experiment.

\begin{figure}[b]
\centerline{\includegraphics[width=0.6\columnwidth]{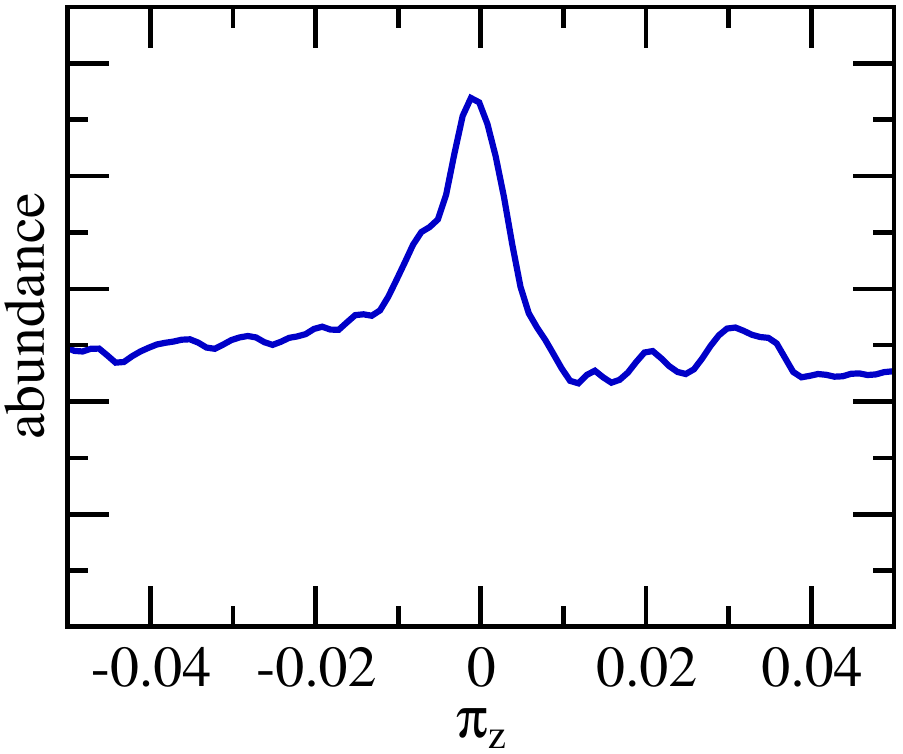}}
\caption{Measured longitudinal momentum distribution for N$_{2}$ molecules ionized with an 780\,nm pulse with duration of 45\,fs.
As in \fig{fig:3200nm} the polarization of the laser pulse is parallel to the extraction field of $F=2.6$\,V/cm.
}
\label{fig:800nm}
\end{figure}
\begin{figure}[b]
\centerline{\includegraphics[width=\columnwidth]{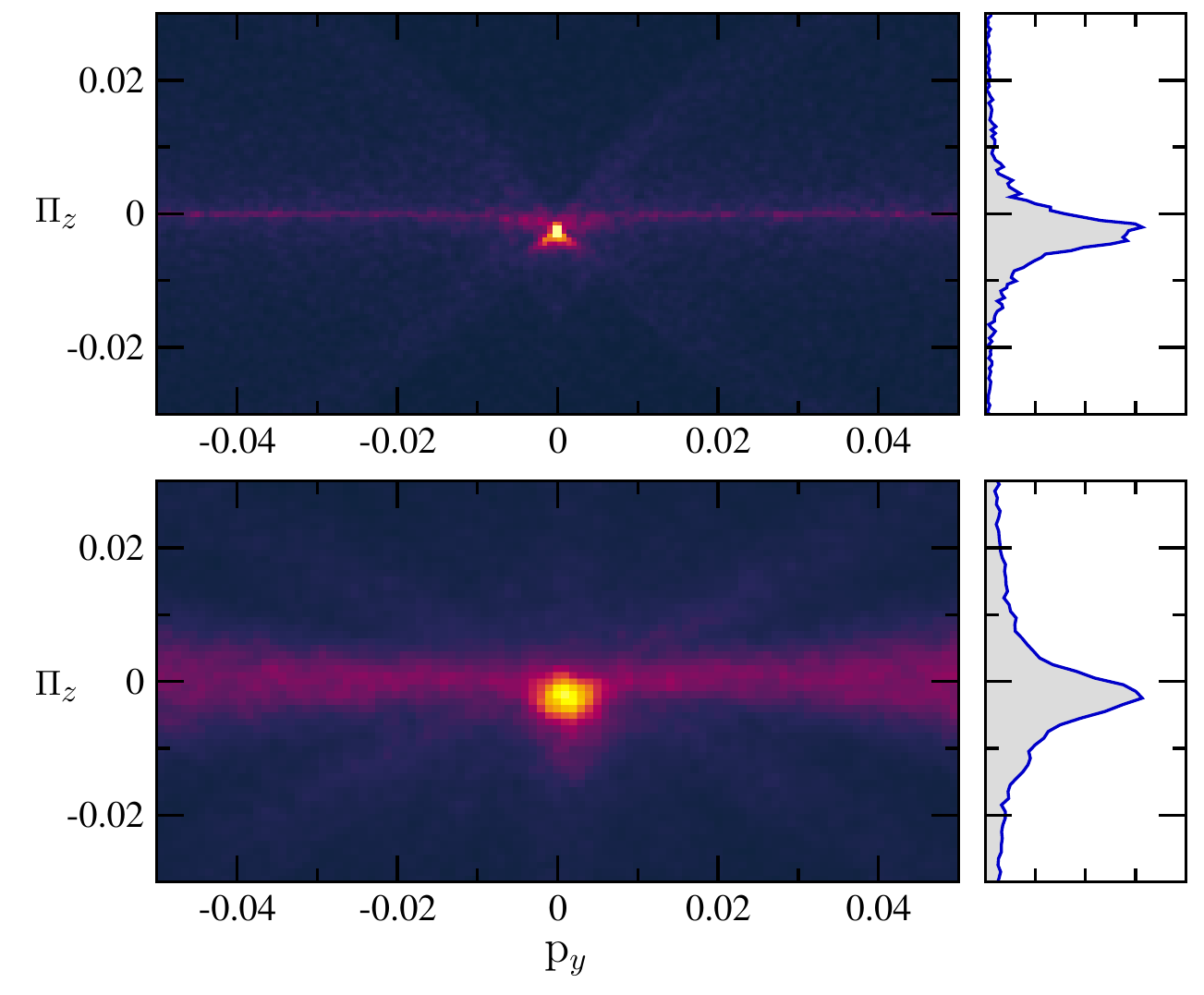}}
\caption{Momentum distribution of photoelectrons with $|\opx|{<}0.01$, whereby the extraction field (2.6\,V/cm) is \emph{perpendicular\/} to the polarisation of the laser pulse (780\,nm), i.\,e.\ $\vec{f}{\perp}\vec{e}_{z}$, which is in contrast to Figs.\ \ref{fig:3200nm} and \ref{fig:800nm}. 
Upper panel: calculation for Ar.
Lower panel: measurement for N$_{2}$.
Each panel shows the 2D distribution and the one integrated over the parallel momentum in the interval $|\opy|\,{<}\,0.005$.}
\label{fig:800nm-perp}
\end{figure}

%
However, a careful inspection of the contour plot in \fig{fig:3200nm}a reveals that the Rydberg peak is asymmetric. 
In fact the peak position $\opeff^{*}$ has a negative value, pointing into the opposite direction of the extraction field $F$, i.\,e., parallel to the actual electric field. This is confirmed by our experiment shown in \fig{fig:800nm}, where for a much shorter wavelength of the driving laser but with the same extraction field a similar peak in the momentum distribution is observed, shifted from zero.
Switching to a setup, where the laser polarization is along the $y$-axis, i.\,e., perpendicular to the extraction field along $\oz$, finally confirms that the peak and its negative offset in $\opeff$ is exclusively due to the extraction field $F$, see \fig{fig:800nm-perp}. Both the calculation and the experiment reveal the peak close to zero and again with a negative shift in the static field direction, here $\opeff_{\perp}$. We also note the qualitative agreement between theory (for argon atoms) and experiment (for $N_{2}$ molecules) which demonstrates that the laser pulse only prepares the initial Rydberg distribution for the ensuing Stark-field dynamics. It takes place far away from the (ionic) core whose exact nature hardly matters.

Backed by these results we are now in a position to concentrate on the Hamiltonian of \eq{eq:ham1} classically.
In a first step we exploit its scaling properties with scaled phase-space variables  
\begin{subequations}\begin{align}\label{eq:var}
\{\sx,\sy,\sz,\srabs\} &=  F^{1/2}\{\ox,\oy,\oz,\orabs\}
\\
\{\spx,\spy,\spz\} &=  F^{-1/4}\{\opx,\opy,\opz\}
\end{align}\end{subequations}
and scaled time $\st=F^{3/4}\ot$,
which eliminate the dependence on $F$ in \eq{eq:ham1} with the new Hamiltonian
\begin{equation}\label{eq:scaledham}
\sH = \frac{\svp^{2}}{2}-\frac{1}{\srabs} - \sz.
\end{equation}
This has two advantages: Firstly, the (experimental) dependence of certain observables on the field strength $F$ can be immediately predicted, e.\,g., $\opz=F^{1/4}\spz$, which applies to all momenta including the effective $\opeff\,{=}\,F^{1/4}\speff$
with $\speff\,{=}\,\spz-\st$.
Secondly, the numerical calculation is largely facilitated since it is reduced to the problem of a single effective field strength ($F{=}1$) with a reasonable (effectively an atomic-scale) range for the phase-space variables. 

\begin{figure}[tb]
\centerline{\includegraphics[width=0.9\columnwidth]{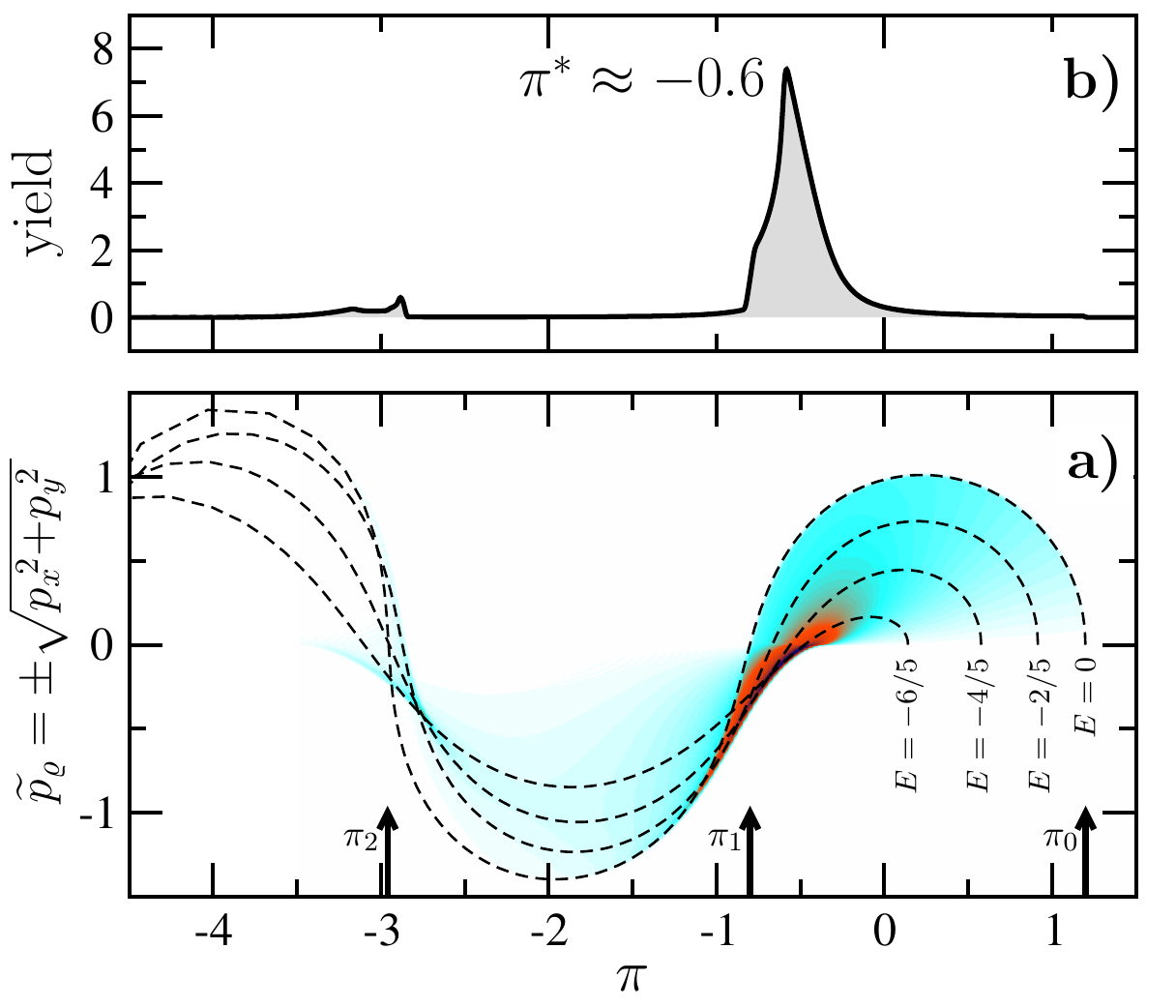}}
\caption{Classical momentum spectra generated from a uniform population of energies $E=-2\ldots0$ and an isotropic distribution of $\cos\theta_{\rm init}$.
{\bf a)} Contour plot as a function of the scaled final momenta $\speff$ and $\widetilde\spr$. The dashed lines show the final momenta for 4 energies indicated in the graph and initial angles in the range $\theta_{\rm init}=0\ldots\arccos(E^{2}/2{-}1)$.
The arrows indicate the location of the zeros $\widetilde\spr(\speff_{n}){=}0$ for $E=0$ according to \eq{final}. 
{\bf b)} Spectrum as a function of $\speff$ after integration over the radial momentum $|\widetilde\spr|\le0.1$.
}
\label{fig:stark-theory}
\end{figure}
\begin{figure}[t!]
\centerline{\includegraphics[width=0.8\columnwidth]{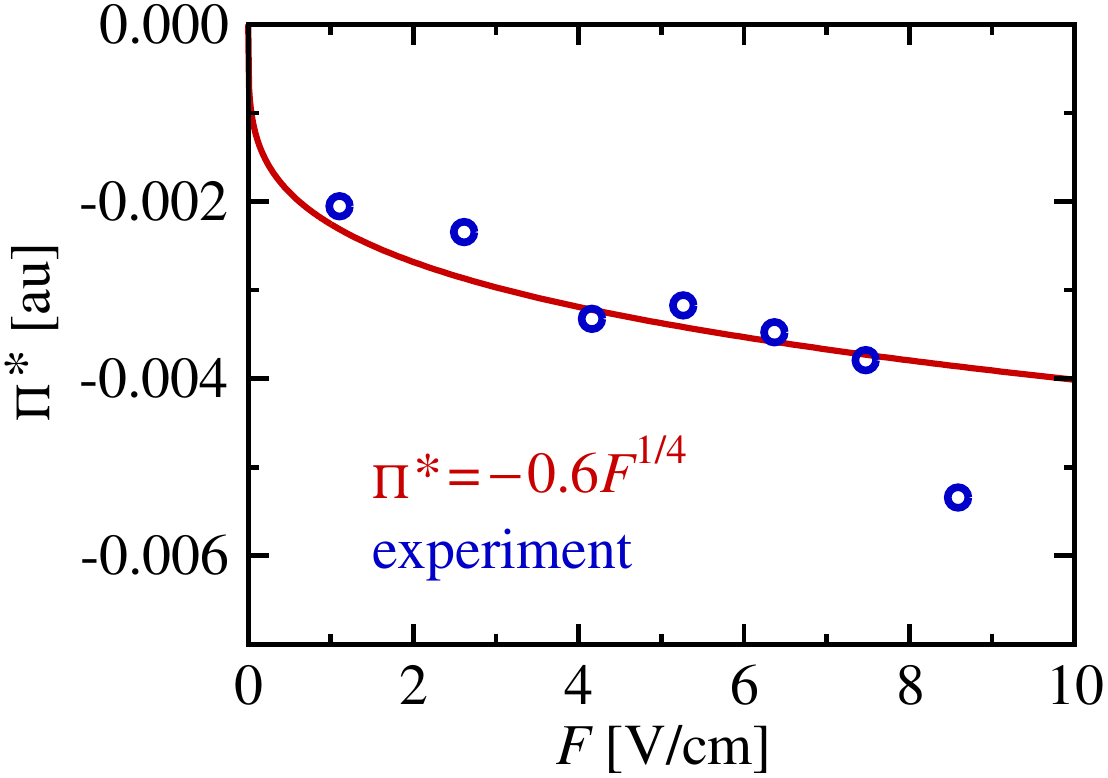}}
\caption{Shifting of peak position $\opeff^{*}$ with the extraction field strength $F$ for photo-ionization of N$_{2}$, cf.\ \fig{fig:800nm}.
Note that the pre-factor in $\opeff^{*}=-0.6F^{1/4}$ has been obtained theoretically from \fig{fig:stark-theory} and is \emph{not} a fitting factor. 
}
\label{fig:fscaling}
\end{figure}

%
From the scaling we immediately deduce that the offset of the peak position $\opeff^{*}$ from zero should scale with $F^{1/4}$.
Indeed, the experiment confirms this dependence as shown in \fig{fig:fscaling}.
While the shift of the peak position with increasing extraction voltage or extraction field, respectively, might have been expected, the very existence of a peak is not obvious on a first glance since one would expect that
the (almost homogeneously) distributed Ryd\-berg population is set free by the extraction field in a certain interval $\Delta E\ldots0$, with the lower bound $\Delta E$ given by the barrier formed by the static field. This should lead to a finite value of the photo-electron spectrum at zero energy but not to a peak, even shifted from zero in the observable $\opeff^{*}$.

To gain insight into the origin of the peak and its position $\speff^{*}$ we take a closer look at the classical dynamics of the Stark Hamiltonian \eqref{eq:scaledham}. 
To this end we propagate trajectories starting at the origin $\sz=\widetilde\sr=0$, with the ``directed'' distance $\widetilde\sr\equiv\pm\sqrt{x^{2}{+}y^{2}}$ from the $z$-axis. 
The sign is positive (negative) if the electron has passed the $z$-axis an even (odd) number of times \cite{angle}.
The initial distribution is uniform in energy $E$ and angle $\cos\theta_{\rm init}$ in accordance with the uniform occupation of Rydberg states. 
It turns out (as most easily seen in semi-parabolic coordinates, discussed below and in the supplement \cite{supp}), that for $E<0$ initial angles $\cos\theta_{\rm init}<E^{2}/2{-}1$ lead to bound trajectories, despite the fact that they have an energy above the barrier $E_{\rm b}=-2$. These are the well-known bound states in the continuum \cite{ga05a}, which are a consequence of the separability of the hydrogen problem in an electric field \cite{ep16,laru11} and account for 1/3 of all trajectories in the range $E=-2\ldots0$. The other 2/3 of the trajectories will escape.
The abundance of the resulting final momenta $\speff$ and $\widetilde\spr$ for those free trajectories is shown as contour plot in \fig{fig:stark-theory}a.
The final momentum $\widetilde\spr=\pm\sqrt{\spx{\!}^{2}+\spy{\!}^{2}}$ perpendicular to the extraction field  oscillates as a function of $\speff$, with decreasing amplitude and period for each fixed energy $E$ (dashed lines). Amplitudes also decrease at a fixed
momentum $\widetilde\spr$  for decreasing energy $E$. 
This topology gives rise to a caustic of final momenta around $\{\speff^{*},\widetilde\spr^{*}\}=\{-0.6,0\}$.

The peculiar structure of the momentum map in \fig{fig:stark-theory}a becomes understandable when taking advantage of the separability of the Stark Hamiltonian in semi-parabolic coordinates \cite{ep16,laru11} 
$\{u,v\}\equiv\{\sqrt{\srabs{+}\sz},\pm\sqrt{\srabs{-}\sz}\}$. 
Here the sign for $v$ corresponds to the sign of $\widetilde\sr$.
In these coordinates $H$ from \eq{eq:scaledham} assumes the form \cite{supp}
\begin{subequations}\label{eq:parabolic}
\begin{align}
\label{eq:uparabolic}
p_{u}{\!}^{2}/2 -Eu^{2}-u^{4}/2 &= 1+\beta,
\\
\label{eq:vparabolic}
p_{v}{\!}^{2}/2 -Ev^{2}+v^{4}/2 &= 1-\beta,
\end{align}
\end{subequations}
where $\beta$ is the separation constant due to the additional dynamical symmetry.
This separation constant is fixed by the initial angle (in real space) according to $\beta=\cos\theta_{\rm init}$ as shown in the supplement \cite{supp}.
The separation \eqref{eq:parabolic} is connected with a new time $\tau$ which is related to the real time $t$  via
$\mathrm{d}t = \big(u^{2}{+}\,v^{2}\big)\,\mathrm{d}\tau$.


While the separability of the dynamics in semi-parabolic coordinates is of no significant help to 
produce the momentum map in \fig{fig:stark-theory}a, it does allow one to understand the origin of the oscillating final electron momentum $\speff(\widetilde\spr)$. First we note \cite{supp} that asymptotically for large $\sz$ or $u$
the cartesian momenta are mapped as $\{\spz,\widetilde\spr\}\to\{u,v\}$. As can be seen from \eq{eq:parabolic}, the dynamics 
is that of a bound ($v)$ and inverted $(u)$ quartic oscillator at an energy $1\mp\beta$.  
The partition of the effective energy between the two modes is controlled by $\beta$. For $\beta =1$ or $\cos\theta_{\rm init} = 0$ all energy is in $u$ leading to the quickest possible escape with the largest value $\speff$ for $u\to\infty$ while  no energy is left for the bound oscillator which stays at the fixed point
$v=0$. Decreasing $\beta$, keeping $E$ fixed, reduces the effective energy in $u$ and provides in turn some energy for oscillations in $v$. Once there is enough energy for the $v$ oscillator to complete half an oscillation, while the electron escapes, it will end up again at $v(\tau{\to}\tau_{\infty})=0$. 
Hereby $\tau_{\infty}$ is the escape time in $u$-direction, which is finite due to the quartic term in \eqref{eq:uparabolic}. 
Thus we can define a series of $\speff_{n}$, for which asymptotically $v=0$ and therefore $\widetilde\spr=0$, with $n=0,1,\ldots$ the number of half oscillations in  $v$. That these values $\speff_{n}$ are finite despite the ever increasing momentum due to the acceleration from the electric field is the consequence of the definition of $\speff=\spz-\st$, cf.\,\eq{eq:peff}, which compensates exactly for this acceleration.
 
Analytical quadratures for the phase space variables of the quartic oscillators in form of  combinations of elliptic functions exist \cite{laru11}, but do not yield simple expressions describing the prominent features of \fig{fig:stark-theory}.
Yet, for the highest energy $E=0$ these integrals simplify and therefore it is possible to determine the zeros of $\widetilde\spr(\speff)$ analytically,
\begin{equation}\label{final}
\speff_{n}=\speff\big|_{\widetilde\spr=0} = \frac{2(1-2n^{2})}{(1+4n^{4})^{1/4}}
\frac{\sqrt{\pi}\Gamma(3/4)}{\Gamma(1/4)},
\end{equation}
as derived in detail in the supplement \cite{supp}. 
The arrows in \fig{fig:stark-theory}b indicate the positions of the $\speff_{n}$ according to \eq{final} in good agreement with the numerical results. 

While the $\speff_{n}$ do not give the exact clustering positions they follow the same qualitative evolution and provide analytical insight for the latter:
The second clustering near $\speff_{2}$ is much weaker because it takes place at
smaller values of $\beta$ as compared to $\speff_{1}$. For the zero crossings at $E=0$ the corresponding values read $\beta_{n}= (1-n^{4})/(1+n^{4})$. Since for escape, i.\,e.\ over-barrier motion, $E^{2}/2 <1+\beta$ 
the phase-space volume contributing to the caustics decreases with the order $n$ and the corresponding peak in the spectrum gets weaker.

The excellent agreement of our classical calculations with our experimental results raises the question why quantum effects do not play a role, in particular given the important role of the potential barrier for the present dynamics. The reason is the weakness of the field which renders any action like quantity very large
as can be directly inferred from the scaling \eq{eq:var}, $S =  {\rm S} F^{-1/4}$. This applies to the
imaginary action of a tunnelling path, so that tunnelling and diffractive phenomena are comparatively weak. The same argument holds true also for the real actions appearing as phases. Since  the action differences 
scale with $F^{-1/4}$ as well and are large for small fields, interference between different trajectories is quenched leaving the classical limit as an excellent approximation. 
This is in contrast to the situation for strong fields where interference patterns are observed \cite{lebo+04, stro+13}, at fields several orders of magnitude stronger than discussed here. 
 
In summary we have shown that the Stark dynamics of escaping Rydberg electrons leads to a peak in the spectrum of the effective momentum $\opeff$ or the time of flight as the experimental observable, changing its position with the field strength according to $\opeff^{*}(F) = -0.6F^{1/4}$. 
This explains the occurrence of the  ``zero-energy structure'' \cite{duca+13}.
The universality of the phenomenon is underlined through the very good agreement with experiments where the initial Rydberg population was created by laser pulses of different wavelengths and for different targets (atoms and molecules). 
This peak is not only an unexpected and intricate consequence of the Stark dynamics, it also has direct implications for the experiment: The peak can be used to gauge the momentum scale if the field is known or vice versa, to determine the field strength with known momentum scale.

This work was supported by the COST Action XLIC (CM 1204) and the Marie Curie Initial Training Network CORINF. 
The experimental work is supported by DFG.


\begin{thebibliography}{10}
\bibitem{angle}
Actually, the motion is independent of the azimutal angle $\phi$. However, if
  the particle crosses the symmetry axis, the angle $\phi$ jumps by 180$^{\circ}$. This
  is taken into account by the prefactor ``$\pm$''.
\end{thebibliography}
\end{document}